\documentclass[%
 reprint,
superscriptaddress,
 amsmath,amssymb,
 aps,
]{revtex4-2}

\usepackage{graphicx}
\usepackage{dcolumn}
\usepackage{bm}
\usepackage{comment}
\usepackage{braket}
\usepackage{upgreek}
\usepackage{appendix}
\usepackage{xcolor}

\begin{document}

\preprint{APS/123-QED}

\title{Poincaré sphere symmetries in four-wave mixing with orbital angular momentum}

\author{M. R. L. da Motta}
\email{mateus.motta@ufpe.br}
\affiliation{Departamento de Física, Universidade Federal de Pernambuco, 50670-901, Recife, Pernambuco, Brazil}
\author{G. B. Alves}
\email{gbie@id.uff.br}
\affiliation{Instituto de Física, Universidade Federal Fluminense, 24210-346, Niterói, Rio de Janeiro, Brazil}
\author{A. Z. Khoury}
\email{azkhoury@id.uff.br}
\affiliation{Instituto de Física, Universidade Federal Fluminense, 24210-346, Niterói, Rio de Janeiro, Brazil}
\author{S. S. Vianna}
\email{sandra.vianna@ufpe.br}
\affiliation{Departamento de Física, Universidade Federal de Pernambuco, 50670-901, Recife, Pernambuco, Brazil}




\date{\today}

\begin{abstract}
We explore a degenerate four-wave mixing process induced by transversely structured light beams in a rubidium vapor cell.
In particular, we consider the nonlinear interaction driven by optical modes contained in the orbital angular momentum Poincaré sphere, which can be parametrized in terms of a polar and an azimuthal angle. In this context we investigate the transfer of spatial structure to two distinct four-wave mixing signals, possessing different propagation directions in space.
We show that under usual assumptions, the output fields can also be described by modes belonging to Poincaré spheres, and that the angles describing the input and output modes are related according to well-defined rules.
Our experimental results show good agreement with the calculations, which predict intricate field structures and a transition of the FWM transverse profile between the near- and far-field regions.
\end{abstract}

\maketitle

{
\section{Introduction}

In recent years, the spatial structure of light has seen a significant increase in research interest, both in fundamental studies, and in applications and technological developments \cite{rubinsztein2016roadmap}.
The understanding of effects attributed to the transverse structure of light in optical phenomena, and the ability to control the spatial degrees of freedom of the light field have allowed numerous advances in the optical sciences \cite{forbes2021structured}. We may highlight fundamental properties of electromagnetic radiation \cite{calvo2005,Ballentine2016}}, quantum optics \cite{gao2018,nape2023quantum}, manipulation of matter \cite{franke2007optical,melo2020optical}, holography \cite{Mandal2022}, information multiplexing \cite{Wilner2015,Gong2019,Gregg2019}, quantum communication \cite{Zhou:15,structured2021}, metrology \cite{Toppel_2014}, and nonlinear light-matter interactions \cite{Faccio2013,Li2013,Pereira2017,Xiong2017,Buono:2018,sonja2021}.

The starting point of these advances can be traced back to the year of 1992, when the seminal work of Allen \textit{et al.} \cite{allen92} established the connection between the orbital angular momentum (OAM) of a light beam and its spatial distribution.
This breakthrough originated the field of light OAM, which over the past three decades has grown immensely, and transformed in such a way as to be recognized today as the more general field of structured light \cite{rubinsztein2016roadmap}.
Shortly after these advancements, the investigation of the role played by OAM in nonlinear optical processes started in second harmonic generation \cite{Dholakia:1996,Courtial:1997,Berzanskis:1998}.
Today, SHG and other second-order optical phenomena offer a highly versatile platform to study the transverse degrees of freedom of light.
Four-wave mixing, a third-order nonlinear optical process, has also been extensively employed in the context of structured light \cite{boyer2008entangled,walker2012trans,akulshin2015distinguishing,akulshin2016arithmetic,offer2018spiral,chopinaud2018high,prajapati2019optical,sonja2021}.

Within the seemingly endless sea of structured light \cite{forbes2021structured}, one finds the optical modes belonging to the so-called OAM Poincaré sphere (PS) \cite{padgett1999poincare}, named in a analogy with the polarization Poincaré sphere. They are given by combinations of Laguerre-Gaussian (LG) modes with topological charges of equal magnitude and opposite handedness, and can be parameterized in terms of polar and azimuthal angles on the sphere. OAM PS modes have been widely employed in three-wave mixing in nonlinear crystals \cite{Faccio2013,Li2013,Pereira2017,Rodrigues:18,rodrigues2022generalized}. The interplay between spin and orbital angular momentum in second harmonic generation has also been demonstrated \cite{Xiong2017,Buono:2018,Pinheiro2022}.

In this work we experimentally investigate the nonlinear wave mixing induced by OAM PS beams in a heated sample of rubidium atoms, and the underlying rules that dictate the transfer of optical spatial structure.
In particular, we consider a degenerate forward four-wave mixing (FWM) process in a configuration where two distinct signals are generated, and we study both of these signals \cite{motta2022spatial,da2023combinations}.
We theoretically describe both generated beams and show that, under the usual set of assumptions, they can also be represented as optical modes contained in Poincaré spheres.
An interesting point is that, as we shall demonstrate, the PS components of both four-wave mixing outputs satisfy selection rules similar to those verified in three-wave mixing processes, where a specular reflection symmetry in the Poincaré sphere is observed in the down conversion process \cite{coutinho2007,Rodrigues:18,rodrigues2022generalized} and the generation of a radial mode spectrum has been demonstrated in second harmonic generation \cite{Pereira2017,Buono2020}.
Interestingly, our FWM setup displays simultaneously both the Poincaré sphere symmetry of one of the generated fields and the appearance of a radial mode spectrum in the spatial structure of the other field generated in the process.
The predicted FWM intensity, as well as the consequences of the symmetry properties, are in good agreement with our experimental results.

\section{Experimental configuration}

\begin{figure}[b]
    \centering
    \includegraphics[width=0.48\textwidth]{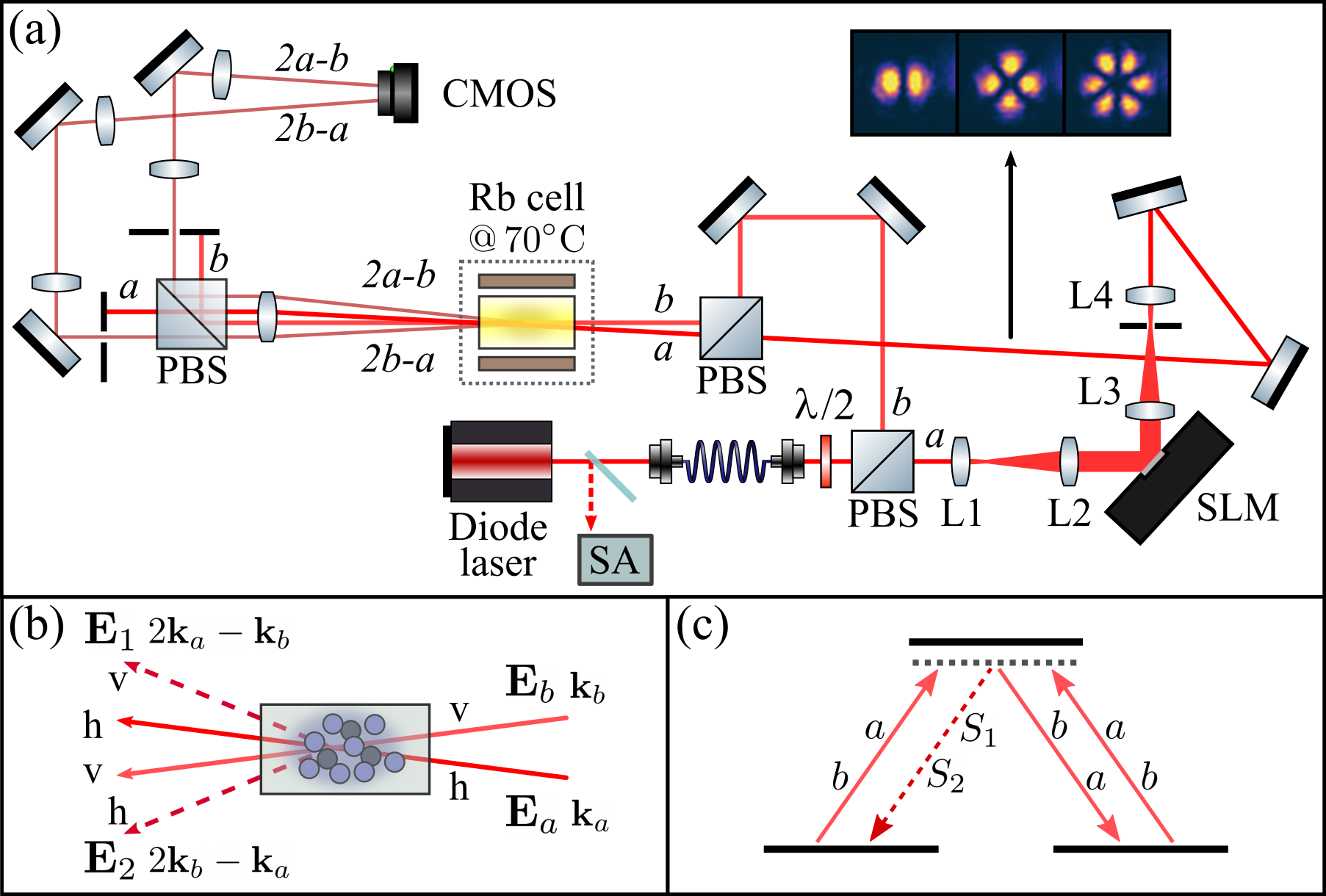}
    \caption{(a) Simplified scheme of the experimental setup for the detection of FWM beam profiles. The pairs of lenses L1-L2 and L3-L4 form telescopes to increase and decrease the diameter of the beam modulated by the SLM. (b) Spatial orientation and polarizations of incident and generated signals near the vapor cell. (c) Generation of the FWM signals $S_1$ and $S_2$ in a three-level system.}
    \label{fig:exp_setup}
\end{figure}

A simplified scheme of our experimental setup is shown in Fig. \ref{fig:exp_setup}. We use a tunable diode laser from Sanyo, model DL7140-201S, with homemade electronics for current and temperature control. A small portion of the laser power goes to a saturated absorption (SA) setup to allow for frequency reference, and then it is coupled to a single-mode fiber to correct the initial transverse profile, which is fairly non-Gaussian.
At the fiber exit, the beam is split in two by a polarizing beam splitter (PBS). We name the transmitted and reflected beams $\mathbf{E}_a$ and $\mathbf{E}_b$, respectively. Beam $\mathbf{E}_a$ is modulated by a spatial light-modulator (SLM) before being sent to the vapor cell, and it carries the non-trivial optical mode. Our SLM is a liquid crystal on silicon (LCOS) SLM from Hamamatsu Photonics, model X10468-02.
Beam $\mathbf{E}_b$ is sent directly to the Rb vapor cell to intersect beam $\mathbf{E}_a$.

The two beams $\mathbf{E}_a$ and $\mathbf{E}_b$, with wave-vectors $\mathbf{k}_a$ and $\mathbf{k}_b$, respectively, and orthogonal and linear polarizations, co-propagate with a small angle of about 10 mrad inside a 5 cm long cell. We detect two four-wave mixing signals generated in the $2\mathbf{k}_a-\mathbf{k}_b$ and $2\mathbf{k}_b-\mathbf{k}_a$ directions, denominated as $S_1$ and $S_2$, respectively.
Since $\mathbf{E}_a$ and $\mathbf{E}_b$ possess orthogonal polarizations, the generated fields $\mathbf{E}_1$ and $\mathbf{E}_2$ are also orthogonally polarized with respect to each other. Moreover, $\mathbf{E}_1$ ($\mathbf{E}_2$) is orthogonally polarized with respect to $\mathbf{E}_a$ ($\mathbf{E}_b$).
This results in a arrangement at the output where the four signals possess alternating polarizations.
Figure \ref{fig:exp_setup}(b) shows the spatial orientation of the incident and generated beams, while Fig. \ref{fig:exp_setup}(c) shows a simplified representation of the nonlinear processes associated with the generation of signals $S_1$ and $S_2$ in a three-level atom.

We arrange the setup in such a way that the waists of both incident beams are located inside the vapor cell, where they intercept. The rubidium cell, containing both ${}^{87}$Rb and ${}^{85}$Rb in natural abundances, is heated to about $70\,^\circ\mathrm{C}$ to increase atomic density. For all measurements, we considered the $\ket{5^2\mathrm{S}_{1/2},F_g=3}\rightarrow\ket{5^2\mathrm{P}_{3/2}}$ transition of ${}^{85}$Rb.
Since the nonlinear output is much weaker than the transmitted beams, scattered light from the input beams was a big problem, especially because we work in a degenerate configuration, and we circumvented it by using spatial filters on the path of both FWM beams.
The waists of the incident beams at the location of the cell were $w_0\approx0.5$ mm. This gives the Rayleigh range value of $z_R=\pi w^2_0/\lambda\approx1$ m. For a cell of length $L=5$ cm, we assume that the thin-medium regime ($L/z_R\ll1$) is always satisfied.

\section{Theory}

The electric fields of the light beams participating in the FWM processes are written as:
\begin{align}    \mathbf{E}_i(\mathbf{r},t)=\frac{1}{2}\boldsymbol{\epsilon}_i\mathcal{E}_i(\mathbf{r}) e^{-i(\mathbf{k}_i\cdot\mathbf{r}-\omega_it)}+\mathrm{c.c.},
\end{align}
$i\in\{a,b,1,2\}$, where $\boldsymbol{\epsilon}_i$ is the polarization direction, $\mathcal{E}_i$ is the slowly varying field amplitude, $\mathbf{k}_i$ is the wave-vector, and $\omega_i=c|\mathbf{k}_i|$ is the frequency. The wave-equations describing evolution of the slowly varying FWM field envelopes $\mathcal{E}_1$ and $\mathcal{E}_2$, as a result of the two  independent FWM processes in directions $(2\mathbf{k}_a-\mathbf{k}_b)$ and $(2\mathbf{k}_b-\mathbf{k}_a)$ are \cite{motta2022spatial}:
\begin{align}
    \left(\frac{i}{2k_1}\nabla^2_\perp+\frac{\partial}{\partial z}\right) \mathcal{E}_1 &= \kappa_1 \mathcal{E}^2_a\mathcal{E}^*_b e^{-i\Delta k_1 z},
    \\
    \left(\frac{i}{2k_2}\nabla^2_\perp+\frac{\partial}{\partial z}\right) \mathcal{E}_2 &= \kappa_2 \mathcal{E}^2_b\mathcal{E}^*_a e^{-i\Delta k_2 z},
\end{align}
where the couplings $\kappa_1$ and $\kappa_2$ are proportional to the nonlinear susceptibilities associated with the two nonlinear processes, $\chi_{2a-b}$ and $\chi_{2b-a}$, $\Delta k_1$ and $\Delta k_2$ are the phase mismatches.
The solutions to these equations can be written as general superpositions of the form:
\begin{align}
    \mathcal{E}_1(\mathbf{r}) &= \sum_{\ell,p} \alpha_{\ell,p}  u_{\ell,p}(\mathbf{r}), \label{eq:E1}
    \\
    \mathcal{E}_2(\mathbf{r}) &= \sum_{\ell,p} \beta_{\ell,p}  u_{\ell,p}(\mathbf{r}), \label{eq:E2}
\end{align}
where $u_{\ell,p}$ is the Laguerre-Gaussian (LG) mode, denoted as:
\begin{align}
    u_{\ell,p}(\mathbf{r}) &= \frac{C_{\ell,p}}{w(z)}\left(\frac{\sqrt{2}r}{w(z)}\right)^{|\ell|} L_{p}^{|\ell|}\left(\frac{2r^2}{w^2(z)}\right) e^{-\frac{r^2}{w^2(z)}} \nonumber
    \\
    &  \times e^{i\ell\gamma}\exp\left(- i\frac{k r^{2}}{2R(z)} + i\Psi_{\mathrm{G}} (z) \right),
\end{align}
where we write the position vector in cylindrical coordinates as $\mathbf{r}=(r,\gamma,z)$, $C_{\ell,p}=\sqrt{2p!/\pi(p+|\ell|)!}$ is a normalization constant, $L_{p}^{|\ell|}(\cdot)$ is the associated Laguerre polynomial, $w(z)=w_0\sqrt{1+(z/z_{R})^{2}}$ is the beam waist, $R(z)=z\left[1+(z_{R}/z)^{2}\right]$ is the curvature radius, $\Psi_{\mathrm{G}}(z)=(N_{\ell,p}+1)\tan^{-1}(z/z_R)$ is the Gouy phase shift, with the total mode order defined as $N_{\ell,p}=2p+|\ell|$.
It is important to note that families of modes $\{u_{\ell,p}\}$ with different spot sizes $w_0$ form different bases.

We consider that the incident fields can be written as $\mathcal{E}_a(\mathbf{r})=\mathcal{E}^0_au_a(\mathbf{r})$ and $\mathcal{E}_b(\mathbf{r})=\mathcal{E}^0_bu_b(\mathbf{r})$, where $\mathcal{E}^0_{a,b}$ gives the total power content of each field, $P_{a,b}=\frac{1}{2}c\varepsilon_0|\mathcal{E}^0_{a,b}|^2$, and $u_{a,b}(\mathbf{r})$ carries their spatial structure.
The coefficients $\alpha_{\ell,p}$ and $\beta_{\ell,p}$ are called the full spatial overlap integrals, and can be expressed as:
\begin{align}
    \alpha_{\ell,p} &= \kappa_1\mathcal{E}^0_1\int^{L/2}_{-L/2}\mathcal{A}_{\ell,p}(z)e^{-i\Delta k_1 z}\mathrm{d}z,
    \\
    \beta_{\ell,p} &= \kappa_2\mathcal{E}^0_2\int^{L/2}_{-L/2}\mathcal{B}_{\ell,p}(z)e^{-i\Delta k_2 z}\mathrm{d}z,
\end{align}
where $\mathcal{E}^0_1=(\mathcal{E}^0_a)^2(\mathcal{E}^0_b)^*$, $\mathcal{E}^0_2=(\mathcal{E}^0_b)^2(\mathcal{E}^0_a)^*$ and
\begin{align}
    \mathcal{A}_{\ell,p}(z)&=\iint u^2_au^*_bu^*_{\ell,p}\mathrm{d}^2\mathbf{r}_\perp, \label{eq:OLA}
    \\
    \mathcal{B}_{\ell,p}(z)&=\iint u^2_bu^*_au^*_{\ell,p}\mathrm{d}^2\mathbf{r}_\perp, \label{eq:OLB}
\end{align}
are the transverse overlap integrals of the product of incident beams on the mode basis with waist $w_0$.
In the thin-medium regime, characterized by $L/z_R\ll1$, only the transverse overlap integrals evaluated at $z=0$ are relevant for calculations \cite{motta2022spatial}.
We may therefore write:
\begin{align}
    \alpha_{\ell,p} &\simeq \kappa_1 \mathcal{E}^0_1 T_1(L) \mathcal{A}_{\ell,p}(0),
    \\
    \beta_{\ell,p} &\simeq \kappa_2 \mathcal{E}^0_2 T_2(L) \mathcal{B}_{\ell,p}(0),
\end{align}
where $T_{j}(L) =\int^{L/2}_{-L/2} e^{-i\Delta k_{j}}\mathrm{d}z=L\,\mathrm{sinc}(\Delta k_j L/2)$, $j=1,2$, can be seen as efficiency measures of the wave mixing processes.
Note that the factors $\kappa_{j}\mathcal{E}^0_{j}T_{j}(L)$ are common for all $(\ell,p)$, and therefore do not affect the mode superpositions of the generated fields. We do not carry these factors further.
Now we can explore scenarios where the incident beams $\mathcal{E}_a$ and $\mathcal{E}_b$ carry different structures.

\begin{figure}[t!]
    \centering
    \includegraphics[scale=0.75]{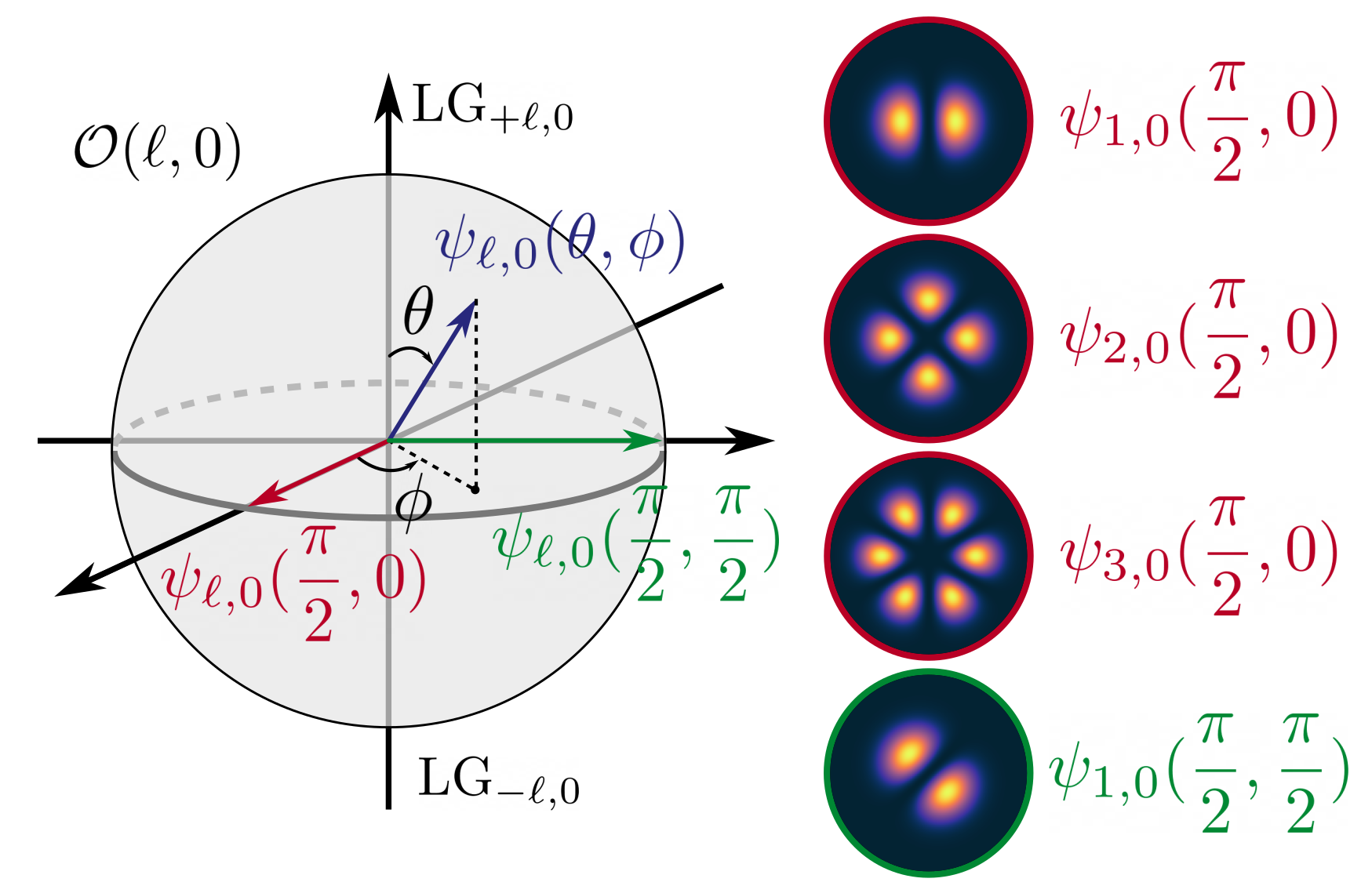}
    \caption{Representation of a spatial mode as a vector parametrized by the polar $\theta$ and azimuthal $\phi$ angles on the OAM Poincaré sphere $\mathcal{O}(\ell,0)$, and intensity profile of different modes on the sphere equator, $\theta=\pi/2$.}
    \label{fig:sphere}
\end{figure}

We will focus on the situation where field $u_b$ is given by a pure Gaussian mode, $u_{0,0}$, and $u_a$ is given by the composition of LG modes contained in the OAM Poincaré sphere $\mathcal{O}(l,0)$ (see Fig. \ref{fig:sphere}):
\begin{align}
    \psi_{l,0}(\theta,\phi) = \cos\frac{\theta}{2}\,u_{l,0}+e^{i\phi}\sin\frac{\theta}{2}\,u_{-l,0}.
\end{align}
Upon substitution in Eqs. (\ref{eq:OLA}) and (\ref{eq:OLB}), we can write the transverse overlap integrals at $z=0$ as:
\begin{align}
    \mathcal{A}_{\ell,p}(0) &= \cos^2\frac{\theta}{2}\Lambda^{ll0\ell}_{000p}+e^{2i\phi}\sin^2\frac{\theta}{2}\Lambda^{-l,-l0\ell}_{000p}+ \label{eq:Alp}
    \\
    &+e^{i\phi}\sin\theta \Lambda^{l,-l0\ell}_{000p}, \nonumber
    \\
    \mathcal{B}_{\ell,p}(0) &= \cos\frac{\theta}{2} \Lambda^{00l\ell}_{000p} + e^{-i\phi}\sin\frac{\theta}{2} \Lambda^{00,-l\ell}_{000p}, \label{eq:Blp}
\end{align}
where
\begin{align}
    \Lambda^{ll^\prime m\ell}_{qq^\prime np} = \iint u_{l,q}u_{l^\prime, q^\prime}u^*_{m,n}u^*_{\ell,p}\big|_{z=0}\mathrm{d}^2\mathbf{r}_\perp,
\end{align}
is the transverse overlap integral of four LG modes with the same waist $w_0$.
The conservation of OAM naturally emerges from the azimuthal integral, restricting the possible values for the topological charges contained in the superpositions for $\mathcal{E}_1$ and $\mathcal{E}_2$. In principle, there is no such restriction on the radial orders, and an infinite number of $p$ modes can contribute to the superpositions of the fields $\mathcal{E}_1$ and $\mathcal{E}_2$.

By substituting Eqs. (\ref{eq:Alp}) and (\ref{eq:Blp}) into Eqs. (\ref{eq:E1}) and (\ref{eq:E2}), we can express both generated fields on the $w_0$ mode basis as:
\begin{align}
    \mathcal{E}_1 &= \sum_p a_{p}\psi_{2l,p}(\vartheta_1,\varphi_1) + e^{i\phi}\sin\theta \, c_p u_{0,p}, \label{eq:E1_rad}
    \\
    \mathcal{E}_2 &= \sum_p b_{p} \psi_{l,p}(\vartheta_2,\varphi_2),
    \label{eq:E2_rad}
\end{align}
where $a_p = \Lambda^{ll0,2l}_{000p}$, $b_p=\Lambda^{00,-ll}_{000p}$, $c_p=\Lambda^{l,-l00}_{000p}$, and the output sphere angles are related to the input angles via:
\begin{align}
    \vartheta_1 &= 2\tan^{-1}\left(\tan^2\frac{\theta}{2}\right),\quad\quad\quad \varphi_1 = 2\phi, \label{eq:angles_1}
    \\
    \vartheta_2 &= \pi-\theta,\quad\quad\quad\quad\quad\quad\quad\quad\,\, \varphi_2 = \phi. \label{eq:angles_2}
\end{align}
The relations given by Eqs. (\ref{eq:angles_1}) and (\ref{eq:angles_2}) are similar to those satisfied in parametric amplification, a second-order nonlinear process \cite{Rodrigues:18,rodrigues2022generalized}. Here we obtained these results because we considered that beam $b$ carried a pure Gaussian distribution.
In this manner, regarding the spatial degrees of freedom, it has a passive role in the wave-mixing process, and we essentially have a single field dictating the transverse mode dynamics. This parallel between three- and four-wave mixing is one aspect that we will emphasize throughout this work.

The FWM fields are generated with a smaller overall size as compared with the input beams, since the generated field amplitudes are dictated by the spatial overlap of the incident modes.
This can be understood intuitively in the case of Gaussian inputs, where we have $\mathcal{E}_{1,2}\sim u^2_{0,0}u^*_{0,0}\sim \exp(-\frac{r^2}{w^2_0/3})$.
By choosing a basis with the appropriate (reduced) minimum waist $\tilde{w}=w_0/\xi$, the number of modes required to represent the FWM fields is reduced.
In fact, for $\xi=\sqrt{3}$, the following restriction on the $p$ orders is established: the sphere modes in $\mathcal{E}_1$ and $\mathcal{E}_2$ are limited to $p=0$ only, while the contribution from the non-vortex modes in $\mathcal{E}_1$ is bound to $0\leq p\leq|l|$. We may then write:
\begin{align}
    \mathcal{E}_1 &= \tilde{a}_0\tilde{\psi}_{2l,0}(\vartheta_1,\varphi_1) + e^{i\phi}\sin\theta\sum^{|l|}_{p=0} \tilde{c}_{p}\tilde{u}_{0,p}, \label{eq:E1_bound}
    \\
    \mathcal{E}_2 &= \tilde{b}_{0} \tilde{\psi}_{l,0}(\vartheta_2,\varphi_2), \label{eq:E2_bound}
\end{align}
where $\tilde{u}$ and $\tilde{\psi}$ are the LG and OAM PS modes with the modified waist $\tilde{w}$, $\tilde{a}_0 = \tilde{\Lambda}^{ll0,2l}_{0000}$, $\tilde{b}_0=\tilde{\Lambda}^{00,-ll}_{0000}$, $\tilde{c}_p=\tilde{\Lambda}^{l,-l00}_{000p}$, where the transverse overlap integral on the modified waist basis is
\begin{align} \label{eq:Lambda_tilde}
    \tilde{\Lambda}^{ll^\prime m \ell}_{qq^\prime n p}(\xi) &= \iint u_{l,q}u_{l^\prime,q^\prime}u^*_{m,n}\tilde{u}^*_{\ell,p}\big|_{z=0}\,\mathrm{d}^2\mathbf{r}_\perp, \nonumber
    \\
    &= \sum_{s} \Lambda^{ll^\prime m\ell}_{qq^\prime ns}\lambda^{\ell}_{s,p}(\xi),
\end{align}
where $\lambda^\ell_{s,p}(\xi)$ are the coefficients for the change of basis $\{u(w_0)\}\rightarrow\{u(w_0/\xi)\}$ \cite{GilOliveira2023}
\begin{align} \label{eq:lambda_cw}
    \lambda^\ell_{s,p}(\xi) = \iint u_{\ell,s}\tilde{u}^*_{\ell,p}\big|_{z=0}\mathrm{d}^2\mathbf{r}_\perp.
\end{align}
We can therefore calculate the mode superpositions $\mathcal{E}_1$ and $\mathcal{E}_2$ on the input waist ($w_0$) basis and modify the coefficients using the second line of Eq. (\ref{eq:Lambda_tilde}) together with Eq. (\ref{eq:lambda_cw}), or perform the calculations directly on the reduced waist ($\tilde{w}$) basis, using the first line of Eq. (\ref{eq:Lambda_tilde}).
In the Appendix \ref{Appendix_A} we give expressions for the relevant overlap integrals, making explicit the radial mode restriction, and in Appendix \ref{Appendix_B} we calculate the change of basis coefficients.

\section{Results and discussion}

\begin{figure*}[ht!]
    \centering
    \includegraphics[width=1\textwidth]{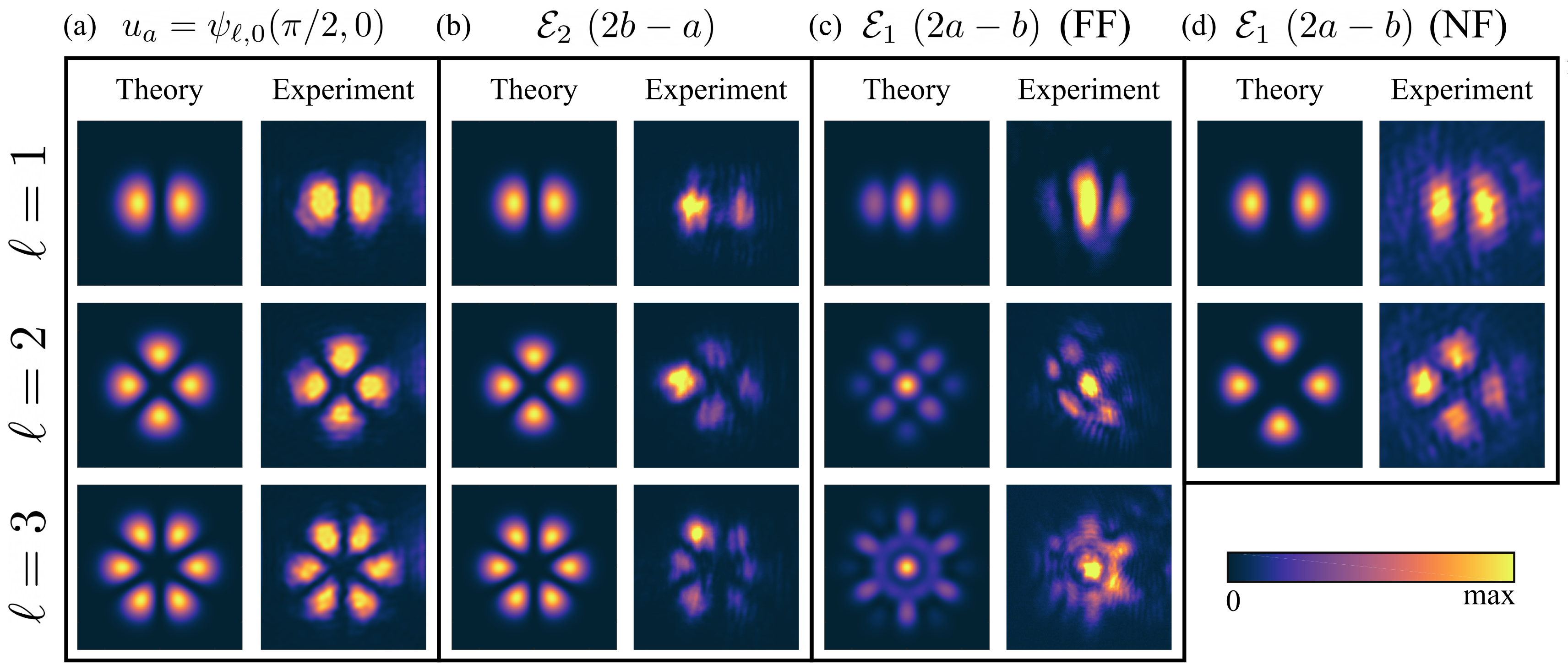}
    \caption{(a) Calculated intensity profiles of modes $\psi_{\ell,0}(\pi/2,0)$, for $\ell=1,2,3$, alongside the corresponding beams prepared in the experiment near the interaction region. Calculated and detected far-field intensity profiles of the FWM signals (b) $S_2$ and (c) $S_1$, resulting from the mixing of $u_a$ given by the modes shown in (a) and $u_b$ given by a Gaussian mode. (d) Calculated and detected near-field intensity profiles of the FWM signal $S_1$ for the cases $\ell=1,2$.}
    \label{fig:FF1_full}
\end{figure*}

First, we performed experiments by setting field $\mathcal{E}_a$ as a mode on the equator ($\theta=\pi/2$) of the PS $\mathcal{O}(\ell,0)$, $\psi^{(a)}_{\ell,0}(\pi/2,0)$ (see Fig. \ref{fig:FF1_full}(a)), and field $\mathcal{E}_b$ as a pure Gaussian beam.
Figures \ref{fig:FF1_full}(b) and \ref{fig:FF1_full}(c) show the calculated and measured far-field intensity profiles of the generated signals, $S_2$ and $S_1$, respectively.
We see that for signal $2a-b$ (Fig. \ref{fig:FF1_full}(c)) we obtain more intricate figures, while the structure of signal $2b-a$ (Fig. \ref{fig:FF1_full}(b)) seems to be dominated by that of the pump in each case.
This is due to the fact that signal $S_1$ has two contributions from the structured pump, and the nonlinear polarization associated with its generation is proportional to $(\psi^{(a)}_{\ell,0})^2$. On the other hand, for $S_2$, which has only one contribution from $\mathcal{E}_a$, the macroscopic polarization is proportional to $(\psi^{(a)}_{\ell,0})^*$.

The central spots present in signal $S_1$ are due to the contribution from 
the $\ell=0$ modes arising from the crossed term in the product $(\psi^{(a)}_{\ell,0})^2$. They only develop in the far-field because of the difference in Gouy phases with respect to the $2\ell$ modes.
This becomes evident when one looks at the near-field intensity distributions of the FWM signal $S_1$ for $\ell=1,2$, shown in Fig. \ref{fig:FF1_full}(d).
This type of transition of the transverse structure has been verified in other situations \cite{Pereira2017,akulshin2016arithmetic}.

\begin{figure}[h!]
    \centering
    \includegraphics[width=0.47\textwidth]{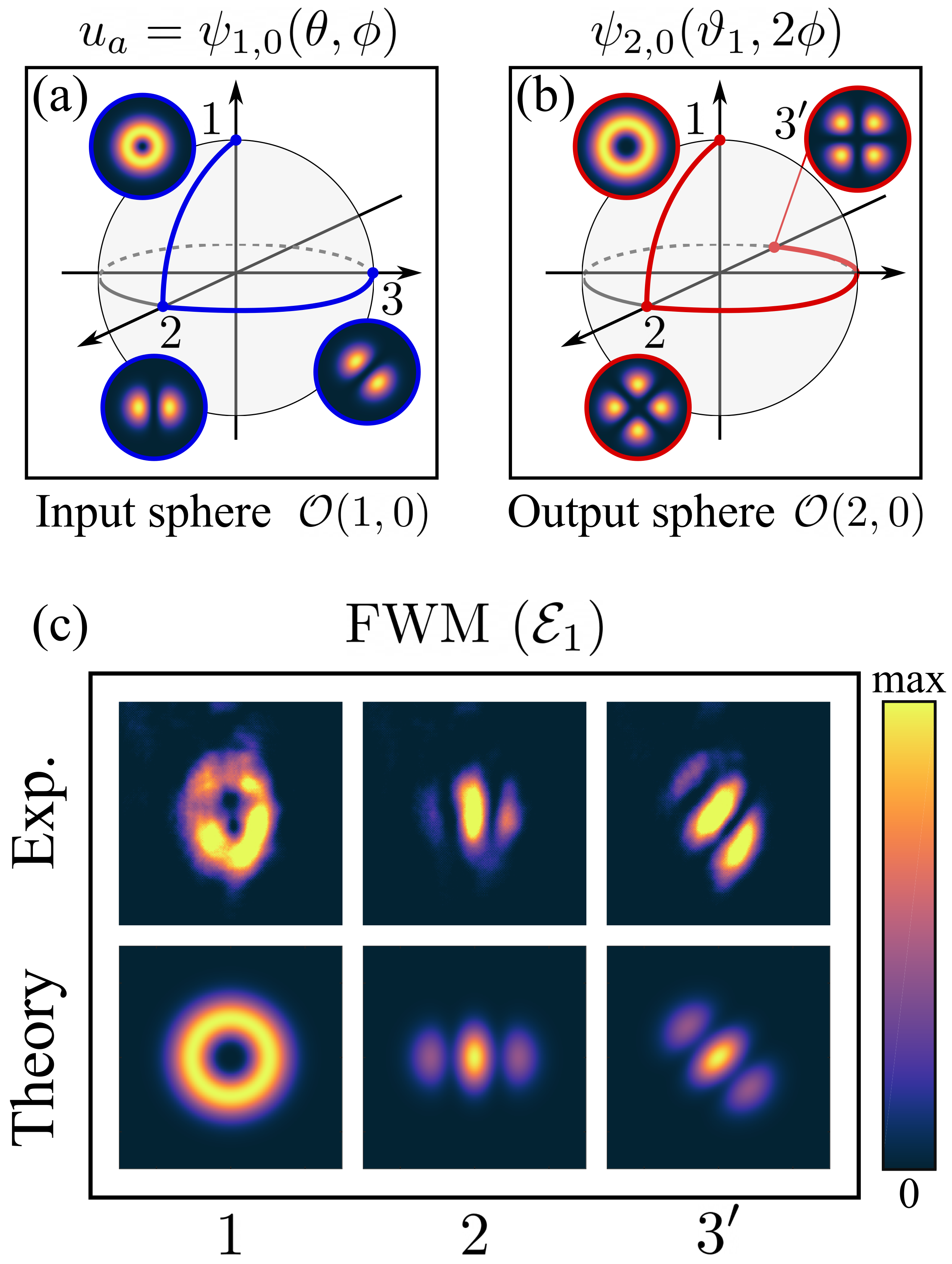}
    \caption{(a) Incident field modes along a path described by points $1,2,3$ on the first order sphere $\mathcal{O}(1,0)$. (b) Corresponding modes on the output sphere $\mathcal{O}(2,0)$ for the FWM signal $\mathcal{E}_1$, when $u_b=u_{0,0}$. Insets show the intensity profiles of the sphere modes on the indicated positions. (c) Measured (top) and calculated (bottom) FWM intensity profiles for signal $\mathcal{E}_1$ when the mode vector of field $u_a$ is located on the points $1,2,3$ along the path shown in (a).}
    \label{fig:path1}
\end{figure}

We also made the pump field $u_a$ occupy different positions on the PS.
First we consider $u_a$ following a path starting on the positive pole, $(\theta,\phi)=(0,0)$, going to $(\pi/2,0)$ on the equator, and then to $(\pi/2,\pi/2)$. We call these points $1,2,$ and $3$, respectively (see Fig. \ref{fig:path1}(a)).
The incident modes in these positions are $\psi_{1,0}(0,0)=u_{1,0}$, $\psi_{1,0}(\pi/2,0)=(u_{1,0}+u_{-1,0})/\sqrt{2}$, and $\psi_{1,0}(\pi/2,\pi/2)=(u_{1,0}+iu_{-1,0})/\sqrt{2}$.
Figure \ref{fig:path1}(b) shows the corresponding path followed by the PS component of the FWM field $\mathcal{E}_1$ on the output sphere $\mathcal{O}(2,0)$.
In Fig. \ref{fig:path1}(c), we show the experimental and theoretical far-field intensity profiles of signal $\mathcal{E}_1$ in this case.
We see that the actual FWM intensity profiles differ from those expected solely from a PS mode $\psi_{2,0}$ on the points $2$ and $3$, corresponding to the first term on the r.h.s. of Eq. (\ref{eq:E1_bound}).
This is due to the contribution from the radial modes in the superposition $\mathcal{E}_1$, which becomes maximum when $\theta=\pi/2$, as seen from the second term on the r.h.s. of Eq. (\ref{eq:E1_bound}).

What is remarkable from these results is that the variation of the azimuthal angle $\phi$ on the input sphere results in a rigid rotation of the FWM intensity profile that is equal to the rotation of the intensity profile of $u_a$. 
This is not obvious since, as already mentioned, (i) the azimuthal angle on the output sphere is doubled, $\varphi_1=2\phi$, and (ii) there are radial modes contributing to the FWM field mode structure.
In fact, this net effect is precisely a result of the combination of these two aspects.
To explain this, we first look at the rotation of the intensity profile of the input PS mode $\psi_{l,0}(\theta,\phi)$, $\mathcal{I}_l(\mathbf{r}_\perp;\theta,\phi)=|\psi_{l,0}(\theta,\phi)|^2$, which can be understood when we write
\begin{align} \label{eq:I_psi}
    \mathcal{I}_l(r,\gamma;\theta,\phi)=|u_{l,0}|^2\left\{1+\sin\theta\cos2l\left(\gamma-\frac{\phi}{2l}\right)\right\}.
\end{align}
We see that the angle $\phi$ shifts the origin of the transverse azimuthal coordinate $\gamma$ by $\phi/2l$:
\begin{align}
    \mathcal{I}_l(r,\gamma;\theta,\phi)=\mathcal{I}_l(r,\gamma-\phi/2l;\theta,0),
\end{align}
thus rotating the intensity profile by $-\phi/2l$.
This can be verified by looking at figures \ref{fig:path1}(a) and \ref{fig:path1}(b), where we see the intensity profiles of the modes on the spheres $\mathcal{O}(1,0)$ and $\mathcal{O}(2,0)$ rotate by 45 degrees when the azimuthal angles vary by 90 and 180 degrees, respectively.
The intensity profile of the FWM field $\mathcal{E}_1$, $I_1=|\mathcal{E}_1|^2$, can be found as
\begin{align}
    &I_1(r,\gamma;\theta,\phi) = \sin^2\theta|U_{l}|^2+\nonumber
    \\
    &+|\tilde{a}_0|^2|\tilde{u}_{2l,0}|^2\left\{1+\sin\vartheta_1\cos\left[4l\left(\gamma-\frac{\phi}{2l}\right)\right]\right\}+
    \\
    &+2\sqrt{2}\tilde{a}_0\tilde{V}^{|2l|}_{0}U_{l}\sin\theta\sin\left(\frac{\vartheta_1}{2}+\frac{\pi}{4}\right)\cos\left[2l\left(\gamma-\frac{\phi}{2l}\right)\right], \nonumber
\end{align}
where the LG radial amplitude $\tilde{V}^{|\ell|}_p(r)$ is defined via $\tilde{u}_{\ell,p}(r,\gamma)=\tilde{V}^{|\ell|}_p(r)e^{i\ell\gamma}$, and $U_l=\sum^{|l|}_{p=0}\tilde{c}_p\tilde{u}_{0,p}$ is the term from Eq. (\ref{eq:E1_bound}) containing the superposition of radial modes.
Thus, we see that just like in Eq. (\ref{eq:I_psi}), $I_1$ presents a shift of the transverse azimuthal coordinate by the amount $\phi/2l$, equal to that of the input PS mode $\psi_{l,0}$:
\begin{align}
    I_1(r,\gamma;\theta,\phi)=I_1(r,\gamma-\phi/2l;\theta,0).
\end{align}

\begin{figure}[ht!]
    \centering
    \includegraphics[width=0.49\textwidth]{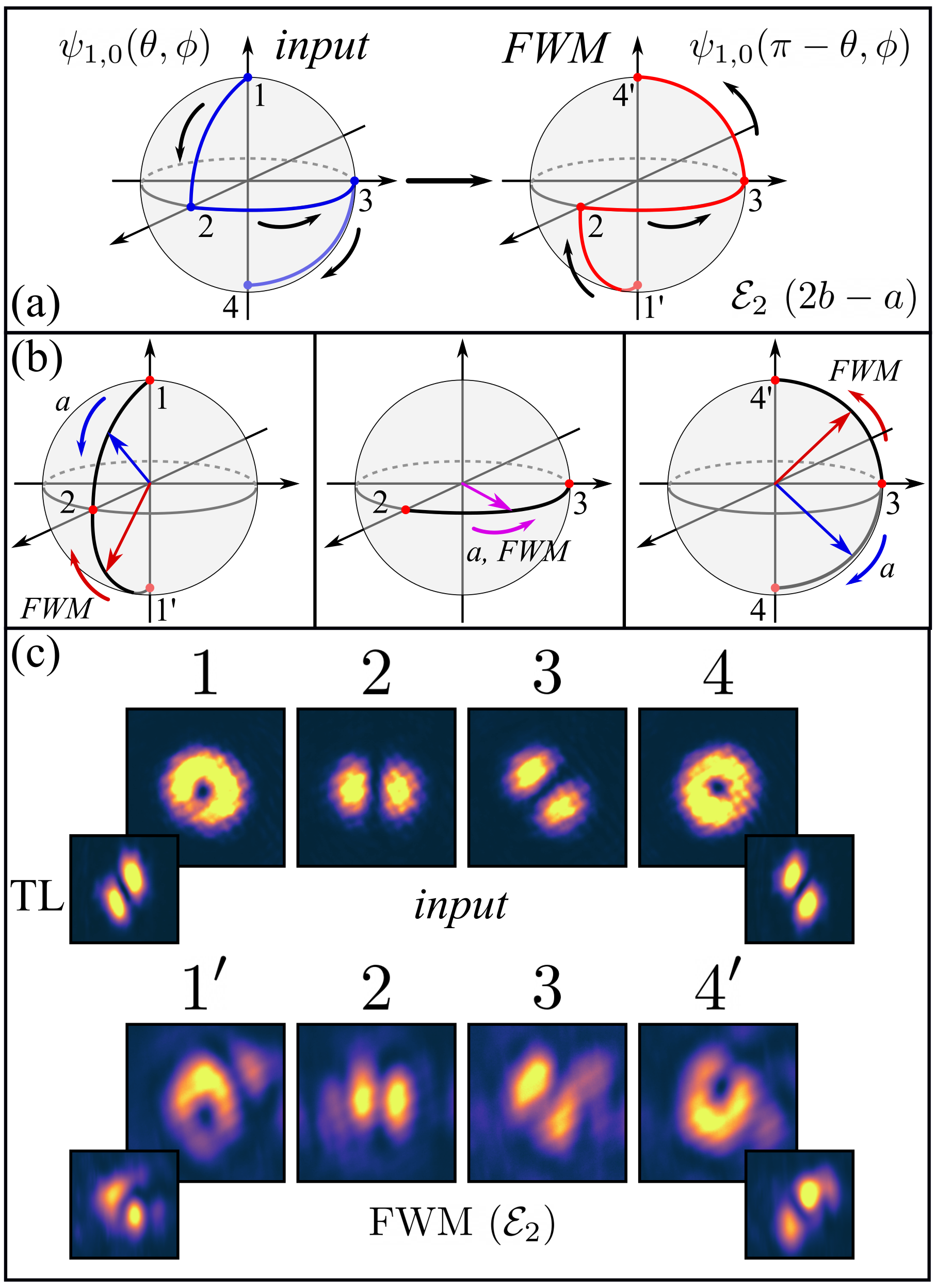}
    \caption{(a) Path followed by the input mode $u_a=\psi_{1,0}(\theta,\phi)$ passing through points $1,2,3,4$ on the sphere $\mathcal{O}(1,0)$, and the corresponding path followed by the FWM signal $\mathcal{E}_2\propto\psi_{1,0}(\pi-\theta,\phi)$, going through points $1^\prime,2,3,4^\prime$. The path on the generated field sphere is a reflection of the path on the input sphere with respect to the equatorial plane. (b) Sections $1,1^\prime-2$, $2-3$, and $3-4,4^\prime$ and the variation of the mode vectors of $u_a$ and $\mathcal{E}_2$ in each one. (d) Detected far-field intensity profiles of the input and FWM signals at the points $1,1^\prime,2,3,4,4^\prime$. For points $1,1^\prime$ and $4,4^\prime$ we also show in the insets the tilted lens (TL) profiles, indicating opposite OAM between input and FWM fields.}
    \label{fig:path_2b_a}
\end{figure}

Next we consider a similar path on the sphere $\mathcal{O}(1,0)$, going through points $1,2,3$ and ending at point $4$, the negative pole, $(\theta,\phi)=(\pi,\pi/2)$.
On this position, the incident mode is $\psi_{1,0}(\pi,\pi/2)=u_{-1,0}$.
In Fig. \ref{fig:path_2b_a}(a) we show the paths followed by $u_a=\psi_{1,0}(\theta,\phi)$ (left), and by the generated field $\mathcal{E}_2\propto\psi_{1,0}(\pi-\theta,\phi)$ (right).
We divide the complete path $1-4$ into three sections, and in Fig. \ref{fig:path_2b_a}(b) we illustrate how the incident and generated mode vectors change on the sphere in each section.
In Fig. \ref{fig:path_2b_a}(c) we show the detected images of the intensity profiles of the incident field $u_a$ at each position (top) and the resulting FWM signal $\mathcal{E}_2$ in each case (bottom).
For the points $1$ and $4$ the insets show the tilted lens (TL) profiles, indicating that in these positions the input and FWM fields possess opposite OAM.
Along the arc $2-3$ the input and generated field modes are degenerate in the sense that the position vector on the first order sphere is the same.
These results indicate the fulfilment of the reflection symmetry for signal $\mathcal{E}_2$.

\section{Conclusions}

In this work, we have investigated the transfer of spatial structures from the input pump fields to the converted fields in a FWM degenerated process in a Rb vapor cell. We have analyzed that the whole interaction can be seen as a two-channel three-wave mixing process, whose spatial structures for the converted fields are independently driven by the square of one input field times the conjugate of the other field. By setting one of the input fields in a fundamental Gaussian mode, we have shown that the spatial structure of each conversion channel, in a thin medium, will be equivalent to what takes place in a second harmonic generation (SHG) and in a parametric down-conversion (PDC) for the fields $\mathcal{E}_1$ and $\mathcal{E}_2$, respectively.

This allowed us to simultaneously explore the symmetries previously reported for these nonlinear processes \cite{Rodrigues:18,rodrigues2022generalized}, by structuring the other input field in a OAM Poincaré sphere $\mathcal{O}(l,0)$. In particular, it was possible to observe the specular reflection symmetry for the field $\mathcal{E}_2$ with respect to the equatorial plane in the OAM sphere. We have also shown that the spatial structure of the other channel ($\mathcal{E}_1$) is a combination of two effects: the generation of radial order modes, and an inner symmetry for the OAM components, similar to what was predicted for a three-wave mixing process \cite{rodrigues2022generalized}. This simultaneous two-channel symmetry of spatial modes can be useful for parallel generation and transmission of correlated fields for quantum information.

\appendix

\section{Radial mode restriction on the reduced waist basis}
\label{Appendix_A}

As mentioned in the main text, with the appropriate choice of a beam waist, it is possible to restrict the number of radial orders contained in the output mode superpositions.
In this Appendix we outline the calculation of the overlap integrals on the reduced beam waist basis, and make this restriction explicit.

The transverse overlap integral on the $\tilde{w}=w/\xi$ mode basis is
\begin{align}
    \tilde{\Lambda}^{ll^\prime m\ell}_{qq^\prime n p}(\xi) = \iint u_{l,q}u_{l^\prime,q^\prime}u^*_{m,n}\tilde{u}^*_{\ell,p}\big|_{z=0}\mathrm{d}^2\mathbf{r}_\perp .
\end{align}
We now look at three different cases, corresponding to the coefficients $\tilde{a}_0,\tilde{b}_0$ and $\tilde{c}_p$ in Eqs. (\ref{eq:E1_bound}) and (\ref{eq:E2_bound}).

Let us focus first on the case $q=q^\prime=m=n=0$. The relevant integral is
\begin{align}
    \tilde{\Lambda}^{ll^\prime 0\ell}_{000 p}(\xi) &= \iint u_{l,0}u_{l^\prime,0}u^*_{0,0}\tilde{u}^*_{\ell,p}\big|_{z=0}\mathrm{d}^2\mathbf{r}_\perp ,\nonumber
    \\
    &=2\pi\delta_{\ell,l+l^\prime} C_{l,0}C_{l^\prime,0}C_{0,0}C_{\ell,p}\xi^{|\ell|+1}\frac{1}{w^4}\times \nonumber
    \\
    &\times\int^\infty_0 (r_w)^{|l|+|l^\prime|+|\ell|} L^{|\ell|}_p(\xi^2r^2_w)e^{-r^2_w(3+\xi^2)/2}r\mathrm{d}r,
\end{align}
where $r_w = \sqrt{2}r/w$.
The choice $\xi=\sqrt{3}$ will be most interesting for us because it allows to establish a maximum value for the possibly coupled $p$ orders.
The only nonzero coefficients are those with $\ell=l+l^\prime$,
\begin{align}
    &\tilde{\Lambda}^{ll^\prime 0,l+l^\prime}_{000 p}(\sqrt{3}) = \frac{8}{\pi w^4}\sqrt{\frac{p!\,3^{1-|l|-|l^\prime|}}{|l|!\,|l^\prime|!\,(p+|l+l^\prime|)!}}\times \nonumber
    \\
    &\times \int^\infty_0(3r^2_w)^{|l+l^\prime|}(3r^2_w)^P L^{|l+l^\prime|}_p(3r^2_w)e^{-3r^2_w}r\mathrm{d}r,
\end{align}
where $P = (|l|+|l^\prime|-|l+l^\prime|)/2$. Then
\begin{align}
    \tilde{\Lambda}^{ll^\prime 0,l+l^\prime}_{000 p}(\sqrt{3}) &= \frac{4}{\pi w^2}\sqrt{\frac{p!\,3^{-1-|l|-|l^\prime|}}{|l|!\,|l^\prime|!\,(p+|l+l^\prime|)!}}\times \nonumber
    \\
    &\times \int^\infty_0 x^{|l+l^\prime|}x^P L^{|l+l^\prime|}_p(x)e^{-x}\mathrm{d}x.
\end{align}

\subsection{Coefficients $\tilde{a}_0$ for the sphere modes contained in $\mathcal{E}_1$}

For $l\cdot l^\prime\geq 0$, $P = 0$, and we can substitute $x^P$ by $L^{|l+l^\prime|}_0(x)=1$. Using the orthogonality relation of the associated Laguerre polynomials, $\int^\infty_0 x^{\alpha}L^{\alpha}_p(x)L^{\alpha}_q(x)e^{-x}\mathrm{d}x = \frac{\Gamma(p+\alpha+1)}{p!}\delta_{p,q}$, we can write
\begin{align}
    \tilde{\Lambda}^{ll^\prime 0,l+l^\prime}_{000 p}(\sqrt{3}) =\left\{
    \begin{array}{cc}
    & 
    \begin{array}{cc}
      \frac{4}{\sqrt{3}\pi w^2}\sqrt{\frac{|l+l^\prime|!}{|l|!\,|l^\prime|!\,3^{|l|+|l^\prime|}}},& \quad \mathrm{for} \quad p = 0, \\
      0,& \quad \mathrm{for} \quad p>0.
    \end{array}
\end{array}\right.
\end{align}
We then see that no radial order $p>0$ is generated.

\subsection{Coefficients $\tilde{c}_p$ for the radial modes contained in $\mathcal{E}_1$}

Now, for $l\cdot l^\prime < 0$, $P=\min(|l|,|l^\prime|)$, and we expand the monomial $x^P$ in terms of Laguerre polynomials as $x^n = n!\sum^n_{j=0} (-1)^j\binom{n+\alpha}{n-j}L^{\alpha}_j(x)=n!\sum^n_{j=0}b^\alpha_{j,n}L^{\alpha}_j(x)$, with $\alpha = |l+l^\prime|$, to write
\begin{align}
    &\tilde{\Lambda}^{ll^\prime 0,l+l^\prime}_{000 p}(\sqrt{3}) = \frac{4}{\pi w^2}\sqrt{\frac{p!\,3^{-1-|l|-|l^\prime|}}{|l|!\,|l^\prime|!\,(p+|l+l^\prime|)!}}P!\times \nonumber
    \\
    &\times \sum^P_{j=0}b^{|l+l^\prime|}_{j,P} \int^\infty_0 x^{|l+l^\prime|}L^{|l+l^\prime|}_j(x)L^{|l+l^\prime|}_p(x)e^{-x}\mathrm{d}x.
\end{align}
The $x$ integral is once again the orthogonality relation of the associated Laguerre polynomials.
Finally, we obtain
\begin{align}
    \tilde{\Lambda}^{ll^\prime 0,l+l^\prime}_{000 p}(\sqrt{3}) &= \frac{4}{\pi w^2}\frac{(-1)^{p}}{(P-p)!}\sqrt{\frac{3^{-1-|l|-|l^\prime|}}{|l|!\,|l^\prime|!\,p!(p+|l+l^\prime|)!}}\times \nonumber
    \\
    &\times P!(P+|l+l^\prime|)!,
\end{align}
for $p\leq P$, and $\tilde{\Lambda}^{ll^\prime 0,l+l^\prime}_{000 p}(\sqrt{3})=0$, for $p>P$.
This result is simplified in the case $l^\prime=-l$, which makes $P=|l|$, and we get
\begin{align}
    \tilde{\Lambda}^{l,-l 00}_{000 p}(\sqrt{3}) &= \left\{\begin{array}{cc}
    & 
    \begin{array}{cc}
       \frac{4}{\sqrt{3}\pi w^2}\frac{(-1)^p}{(|l|-p)!}\frac{|l|!}{p!\,3^{|l|}},& \quad \mathrm{for} \quad p \leq |l|, \\
      0,& \quad \mathrm{for} \quad p>|l|.
    \end{array}
\end{array}\right.
\end{align}

\subsection{Coefficients $\tilde{b}_0$ for the sphere modes contained in $\mathcal{E}_2$}

Next, for $l=l^\prime=q=q^\prime=n=0$, we have
\begin{align}
    \tilde{\Lambda}^{00m\ell}_{000p}(\xi)&=\iint u^2_{0,0}u^*_{m,0}\tilde{u}^*_{\ell,p}\mathrm{d}^2\mathbf{r}_\perp, \nonumber
    \\
    &=2\pi\delta_{\ell,-m}C^2_{0,0}C_{m,0}C_{\ell,0}\xi^{|\ell|+1}\frac{1}{w^4}\times \nonumber
    \\
    &\times\int^\infty_0 (r_w)^{|m|+|\ell|}L^{|\ell|}_p(\xi^2r^2_w)e^{-r^2_w(3+\xi^2)/2}r\mathrm{d}r.
\end{align}
The OAM conservation dictates $\ell=-m$, and thus for $\xi=\sqrt{3}$, we can arrive at the expression
\begin{align}
    \tilde{\Lambda}^{00m,-m}_{000p}(\sqrt{3}) = \frac{4}{\pi w^2}\sqrt{3^{-1-|m|}}\frac{(2|m|)!}{(|m|!)^2}\delta_{p,0}.
\end{align}

\section{Coefficients for changing waist bases}
\label{Appendix_B}

We may expand the mode $u_{\ell,p}$, with waist $w$, on the basis of modes $\tilde{u}_{l,q}$, with waist $\tilde{w}=w/\xi$ as
\begin{align} \label{eq:waist_change_exp}
    u_{\ell,p} = \sum_{l,q}\lambda^{\ell,l}_{p,q}(\xi)\tilde{u}_{l,q},
\end{align}
where the expansion coefficients are
\begin{align}
    \lambda^{\ell,l}_{p,q}(\xi) &= \iint u_{\ell,p}\tilde{u}_{l,q}^*\big|_{z=0}\mathrm{d}^2\mathbf{r}_\perp.
\end{align}
Since we must have $\ell=l$, we drop one of the upper indices, to write
\begin{align} \label{eq:lambda_pql}
    \lambda^{\ell}_{p,q}(\xi)&=\frac{\pi}{2} C_{\ell,p}C_{\ell,q}\xi^{|\ell|+1} \nonumber
    \\
    &\times\int^\infty_0 x^{|\ell|}L^{|\ell|}_p(x)L^{|\ell|}_q(\xi^2x)e^{-x(1+\xi^2)/2}\mathrm{d}x,
\end{align}
where we made the change of variable $x=2r^2/w^2$.
For $\xi=1$, we obtain $\lambda^{\ell}_{p,q}(1)=\delta_{p,q}$, which is expected.
To obtain an analytical expression, we can employ the generating function for the Laguerre polynomials $\sum^\infty_{n=0} t^n L^{\alpha}_{n}(x) = (1-t)^{-(\alpha+1)}e^{-tx/(1-t)}$.
Differentiating $p$ times with respect to $t$, and making $t=0$, we get
\begin{align}
    L^\alpha_p(x) = \frac{1}{p!}\frac{\partial^p}{\partial t^p}\left[\frac{e^{-tx/(1-t)}}{(1-t)^{\alpha+1}}\right]\bigg|_{t=0},
\end{align}
and we can rewrite the integral in (\ref{eq:lambda_pql}) as
\begin{align}
    &\frac{1}{p!q!}\frac{\partial^p}{\partial t^p}\frac{\partial^q}{\partial t^{\prime q}}\frac{1}{\left[(1-t)(1-t^\prime)\right]^{|\ell|+1}}\int^\infty_0 x^{|\ell|}e^{-b(t,t^\prime)x}=\nonumber
    \\
    &=\frac{|\ell|!}{p!q!}\frac{\partial^p}{\partial t^p}\frac{\partial^q}{\partial t^{\prime q}} \frac{1}{\left[b(t,t^\prime)(1-t)(1-t^\prime)\right]^{|\ell|+1}}\Bigg|_{t,t^\prime =0},
\end{align}
with $b(t,t^\prime)=\frac{1+\xi^2}{2}+t/(1-t)+\xi^2t^\prime/(1-t^\prime)$.
By carrying the indicated calculations, one obtains
\begin{align}
    &\lambda^\ell_{p,q}(\xi)=(-1)^p\sqrt{\frac{p!q!}{(|\ell|+p)!(|\ell|+q)!}}\left(\frac{2\xi}{1+\xi^2}\right)^{|\ell|+1}\times \nonumber
    \\
    &\times\sum^p_{n=0}(-1)^n\frac{(q+p+|\ell|-n)!}{n!(p-n)^!(q-n)!}\left(\frac{1-\xi^2}{1+\xi^2}\right)^{q+p-2n}.
\end{align}

\bibliography{apssamp}


\end{document}